\documentclass[jmp]{revtex4-1}

\usepackage{amssymb}
\usepackage{amsmath}
\usepackage{amsfonts}
\usepackage{amsthm}
\usepackage{graphicx}
\usepackage{ulem}
\usepackage[mathscr]{eucal}
\usepackage{bbm}
\usepackage{epsfig}

\def\vp{{ \varphi}}
\def\ec{{ \varepsilon_{c} }}
\def\ee{{\varepsilon}}
\def\ch{{ \chi_{c} }}
\def\ex{{ \mathbf{e}_{1} }}

\def\ez{{ \mathbf{e}_{3} }}

\def\rr{{ \mathbf{r}}}
\def\ro{{ \mathbf{r}_{0}}}
\def\kk{{ \mathbf{k} }}
\def\kx{{ k_{x} }}
\def\kz{{ k_{z} }}

\def\kb{{ k_{b} }}
\def\hb{{ h_{b} }}
\def\dl{{ \delta_{0}}}

\def\cb{{ \mathscr{C} }}
\def\hc{{ \mathscr{H} }}
\def\hcd{{\mathscr{H}_{2}}}
\def\hv{{ \widehat{\text{H}} }}

\def\al{{ \alpha}}
\def\ald{{ \alpha_{2}}}
\def\be{{\beta}}
\def\bed{{\beta_{2}}}

\def\re{{ \text{Re} }}
\def\im{{ \text{Im} }}

\begin{document}


\title{The resonant nonlinear scattering theory
with bound states in the radiation
continuum and the second harmonic generation}

\author{ R\'{e}my F. Ndangali and Sergei V. Shabanov \\ Department of Mathematics, University of Florida, Gainesville, FL 32611, USA}

\begin{abstract}
A nonlinear electromagnetic scattering problem is studied
in the presence of bound states in the radiation continuum.
It is shown that
 the solution is not analytic
in the nonlinear susceptibility and the conventional perturbation
theory fails. A non-perturbative approach is proposed
and applied to the system
of two parallel periodic arrays of dielectric cylinders with a second order nonlinear susceptibility. This scattering system is known to have  bound states in the radiation continuum.
In particular, it is demonstrated that,
for a wide range of values of the nonlinear susceptibility,
the conversion rate of the incident fundamental harmonic into the second one can be as high as 40\% when the distance between the arrays is as low as a half of the incident radiation wavelength.
The effect is solely attributed to the presence of
bound states in the radiation continuum.
\end{abstract}

\maketitle

\section{Introduction}
\label{sec:0}

A conventional approach to nonlinear electromagnetic
scattering problems is based on the power series
expansion in a nonlinear susceptibility $\chi_c$.
For example, for the 2nd order susceptibility,
the physical parameter that determines nonlinear effects
is $\chi_c E_r\ll 1$
where $E_r$ is the electric field at the scattering structure.
The smallness of $\chi_c E_r$ justifies the use
of perturbation theory and the solution is analytic
in $\chi_c$. The situation is different if
the scattering structure has resonances.

Planar periodic structures (e.g., gratings) are known
to exhibit sharp scattering resonances when illuminated
by electromagnetic waves (for a review see, e.g.,
\cite{abajo,b17}). Furthermore,
it is known (see, e.g., \cite{b17}) that in such structures
a local electromagnetic field $E_r$ is amplified
if the structure has narrow resonances:
$E_r\sim E_i/\sqrt{\Gamma}$, where $E_i$ is the amplitude
of the incident wave and $\Gamma$ is the width of the resonance.
Consequently, optical nonlinear effects are amplified
if the system has a sufficiently narrow resonance.
For example, an amplification
of the second
harmonics generation by a single periodic array of dielectric
cylinders \cite{b23} and by other
single-array periodic systems \cite{shg} have
been reported, but no significant flux conversion rate,
comparable to that in conventional methods of second
harmonic generation, has been found.
Owing to the smallness of $\chi_c$ and a {\it finite}
$1/\sqrt{\Gamma}$, the condition $\chi_cE_r\ll 1$
still holds for the studied structures.

If two identical planar periodic structures are aligned parallel
and separated by a distance $2h$, then it can be shown
that for each resonance associated with the single structure,
the combined structure has two close resonances whose
width depends continuously on $h$ so that the width of one of the resonances vanishes, i.e.,
$\Gamma(h)\rightarrow 0$ as $h\rightarrow h_b$ for
some discrete set of distances $h_b$, for sufficiently large $h$~ \cite{svs0,b17,jmp}. This means
that the system has {\it bound states in the radiation continuum}.
Their existence was first predicted  in quantum mechanics
by von Neumann and Wigner \cite{b1} in 1929 and later
they were discovered in some atomic systems \cite{b3}
(see also  \cite{b5,b6,b7} for more theoretical
studies).
Their analog in
Maxwell's theory has only attracted attention recently
\cite{svs0,bsp1,bsp2,bsp3}.
In particular,  for a system of two parallel arrays of periodically
positioned subwavelength dielectric cylinders (depicted
in Panel (a) of Fig.~\ref{fig:1}),
the existence of bound states in the radiation continuum
has been first established in numerical
studies of the system \cite{svs0}.
A complete classification of bound states as well as
their analytic form for this system is given
in \cite{jmp}  for TM polarization. It is also shown \cite{jmp} that bound
states exist in the spectral range in which
more than one diffraction channel are open.
From the physical point of view, bound states in the radiation
continuum are localized solutions of Maxwell's equations
like waveguide modes, but in contrast to the latter their
spectrum lies in the spectrum of scattering radiation (diffraction) modes.

The perturbation theory parameter
$\chi_cE_r\sim
\chi_cE_i/\sqrt{\Gamma(h)}$, as defined above, can no longer be
considered small if bound states
in the radiation continuum are present. This qualitative assessment should be taken with a precaution. In the present study, a
rigorous analysis
of the nonlinear scattering problem by means of the formalism of
Siegert states (appropriately extended to periodic
structures \cite{rs2})
shows that no divergence of a local field
occurs as $h\rightarrow h_b$. However, the conventional
perturbative approach fails because the solution
is not analytic in $\chi_c$.
The situation can be compared with
a simple mechanical analog. Consider a scattering problem
for a particle on a line in a hard core repulsive potential
$V(x)=g/x^2$, $g>0$. No matter how small $g$ is, the particle never
crosses the origin $x=0$ and a full reflection
occurs, but it does so when $g=0$ (a full transmission).
So, the scattering amplitude is not analytic in $g$.
Other, more sophisticated, examples of quantum systems
with such properties are studied
in \cite{klauder}.

The purpose of the present study is twofold.
First, the nonlinear scattering problem is studied
in {\it the presence of bound states in the radiation
continuum}. A non-perturbative approach is developed to solve
the problem.
Second, as an application of the developed
formalism, the problem of the second harmonic generation
is analyzed with an example of the system
depicted in Fig.~\ref{fig:1} (Panel (a)).

\subsection{An overview of the results}

In Section~\ref{sec:1}, the nonlinear resonant
scattering problem with bound states in
the radiation continuum
is transformed into  a system of integral
equations.
A non-perturbative method
is proposed to solve these equations
in the approximation that
 takes into account two nonlinear
effects: a second harmonic generation in the leading
order of $\chi_c$,  and the fundamental harmonic generation
by mixing the second and fundamental harmonics
in the leading order in $\chi_{c}$.
This is the second order effect
in $\chi_c$ known in the theory
as the {\it optical rectification}. The latter is shown to be necessary to ensure the energy flux conservation.

The formalism is illustrated with an example of
two parallel periodic arrays of dielectric cylinders
shown in Panel (a) of Fig.~\ref{fig:1}.
The analysis is based on
the subwavelength approximation (Section~\ref{sec:2})
when the incident wave length is larger
than the radius $R$ of the cylinders. If $k$ is the magnitude
of the wave vector, then the theory has three small
parameters:
$$
\delta_0(k) = {\textstyle\frac 14}\,(kR)^2(\varepsilon_c-1)
\ll 1
\,,\ \ \ \chi_c\ll 1\,,\ \ \ |\Delta h|=|h-h_b|\ll 1
$$
where all the distances are measured in units of
the structure period, in particular $R<1/2$. $\delta_0(k)$
is the scattering phase for a single cylinder,
$\varepsilon_c$ is the linear dielectric susceptibility,
and the amplitude of the incident wave is set to one,
$E_i=1$. With this choice of units, all three parameters
are dimensionless.
The scattering amplitudes of the fundamental and second
harmonics are explicitly found in Section~\ref{sec:3}.

In Section~\ref{sec:4}, it is shown that the ratio of
the flux of the second harmonic
along the normal direction and the incident flux is
$$
\sigma_2 = C \chi_c^{2}E_r^4
$$
where $E_r=E_r(\chi_c,\Delta h,\delta_0)$ is the
fundamental field on the cylinders, and $C=C(\delta_0,\Delta h)$
is some function.
The function $E_r$ is {\it non-analytic} in the
vicinity of zero values of its arguments. The non-perturbative
approach of Section~\ref{sec:1}
is used to prove that
the generated flux of the second harmonic
attains its maximal value when the small parameters
satisfy the condition
\begin{equation}
\label{condition}
(\Delta h)^4\delta_0^3(k_b) = \vartheta\,\chi_c^2
\end{equation}
where $\vartheta\ll 1$ is a numerical constant, and $k_b$ is the magnitude of
the wave vector of the bound state that occurs at $h=h_b$.
Under this condition, $\sigma_2$ becomes
analytic in the scattering phase $\delta_0$ so that in
the leading order,
$$
\sigma_{2,{\rm max}}\approx 4\pi k_b^{-1}\delta_0(k_b)
$$
An interesting feature to note is the independence
of the conversion efficiency on the nonlinear
susceptibility $\chi_c$ (in the leading order
in the scattering phase $\delta_0$). In other words,
given a nonlinear susceptibility $\chi_c$, by a fine
tuning of the distance between the arrays one can always
reach the maximal value which is only determined by
the scattering phase at the wave length of a bound state.
The lowest value of $k_b$ for the system considered
occurs just below the first diffraction threshold
(the wavelength is slightly larger than the structure
period) \cite{jmp}, i.e., $k_b\approx 2\pi$. Taking, for
example, $R=0.15$ and $\varepsilon_c =2$
(so that $\delta_0(2\pi)\approx 0.22$),
the conversion rate reads
$\sigma_{2,{\rm max}} \approx 0.44$, that is,
about $44\%$ of the incident flux is converted
into the second harmonic flux, which is comparable
with the conversion rate achieved in slabs (crystals)
of optical nonlinear materials \cite{crystals}.

 From the physical point of view,
the scattering structure plays the role of a resonator
with the quality factor inversely proportional to $\Gamma$.
The field in the resonator is not uniform and has periodic
peaks of the amplitude $E_r\sim E_i/\sqrt{\Gamma}$ due
to a constructive interference of the scattered
fundamental harmonic. The second
harmonic is produced by the induced dipole radiation
of point scatters located at these peaks. The induced dipole
strength is proportional to $\chi_c E_r^2$. The dipoles
are excited by the incident wave and, due to
their periodic arrangement, they
radiate in phase producing a plane wave in
the asymptotic region (just like a phased array antenna).
If the system has a resonance whose width $\Gamma$ can be continuously driven to zero by changing a physical parameter
of the system, i.e., the system has
a bound state in the radiation continuum, then the strength
of the induced
dipoles radiating the second harmonics can be magnified
as desired, but the resonator cannot be excited by the incident
radiation if $\Gamma=0$ (a bound state is decoupled
from the radiation continuum). So, the optimal width $\Gamma$
at which the second harmonic amplitude is maximal occurs
for some $\Gamma\neq 0$, which explains the existence of
conditions like (\ref{condition}). Since the second harmonic
is generated by point scatterers, the phase matching condition,
needed for optically nonlinear crystals, is not required.
The energy flux of the incident radiation is automatically
redistributed and focused on the scatterers owing to
the constructive interference. Thanks to these physical
features, an active length at which the conversion rate is maximal is close to $2h_b$ whose smallest value for the system
studied is roughly a half of
the wave length of the incident light \cite{svs0,jmp}
(i.e. for an infrared incident radiation it is about a few hundreds
nanometers).

\section{The nonlinear resonant scattering theory}
\label{sec:1}

Suppose that a scattering system has a translational
symmetry along a particular direction and has
non-dispersive linear and
second-order nonlinear dielectric susceptibilities,
$\varepsilon$ and $\chi$, respectively. When the electric field is parallel to the translational symmetry axis (TM Polarization), Maxwell's equations are reduced to
the scalar nonlinear wave equation
\begin{equation}
\frac{1}{c^2}\partial_{t}^{2}\left(\varepsilon E+\frac{\chi}{4\pi}E^2\right)=\Delta E
\label{eq:1}
\end{equation}
Let the coordinate system be set so that the functions
$\varepsilon -1\geq 0$ and $\chi\geq 0$ have support bounded
in the $z-$direction and the system has the translational
symmetry along the $y-$direction. In this case, $\varepsilon$,
$\chi$ and $E$ are functions of $z$ and $x$.
In the asymptotic regions $|z|\rightarrow \infty$,
Eq. (\ref{eq:1}) becomes a linear wave equation.
So, the scattering problem can be considered for
a plane wave of the frequency $\omega$  that propagates
from the asymptotic region
$z\rightarrow -\infty$ to the region $z\rightarrow\infty$.
Furthermore, it is assumed that the functions
$\varepsilon$ and $\chi$ are piecewise constant, i.e.,
$\varepsilon =\varepsilon_c=const$ and
$\chi=\chi_c=const$ in regions occupied by the scattering
system. A conventional treatment of the problem
is based on the assumption that the solution $E$ is
 analytic in $\ch$ and, therefore, can be represented as
a power series expansion,
\begin{equation}
E=2 \text{Re}\left\{ E_{1} e^{-i\omega t}+\ch E_{2}e^{-2i \omega t}+\chi_{c}^{2}\left(E_{3,1}e^{-i\omega t}+E_{3,3}e^{-3i\omega t}\right)+\ldots\right\}
\label{eq:2}
\end{equation}
where $E_{1}$ is the amplitude of the fundamental harmonics in the zero order of $\ch$, $E_{2}$ is the amplitude of the second harmonics in the first order of $\ch$, and so on. This assumption is not true if the system has bound states in the radiation continuum. Indeed, a general solution has the form $E=E_{L}+E_{NL}$,
where $E_L$ is the solution when $\chi_c=0$ and $E_{NL}$ is the correction due to nonlinear effects. Let $\chi$ be written as $\chi=\chi_{c}\eta$, where $\eta$ is the indicator function of the region occupied by the scattering system, i.e., its value is 1 in that region and 0 elsewhere. Then, if $\widehat{G}$ is the Green's function of the operator $\frac{\ee}{c^2}\partial_{t}^{2}-\Delta$ with appropriate (scattering)
boundary conditions, the function $E_{NL}$ satisfies the integral equation
\begin{equation*}
E_{NL}=-\frac{\ch}{4\pi c^2}\widehat{G}\left[\eta\partial_{t}^{2}(E_{L}+E_{NL})^{2}\right]
\label{eq:2.1}
\end{equation*}
The power series
expansion (\ref{eq:2})
can be obtained by the method of successive approximations
for this integral equation, provided the series
is proved to converge.
According to scattering theory \cite{b2,b4}, the
Fourier transform of  $\widehat{G}$ is meromorphic in $k^2=\frac{\omega^{2}}{c^2}$. As is clarified shortly (see discussion of Eq.(\ref{eq:12})), its real poles correspond to bound states in the radiation continuum. Hence, in the presence of a real pole $k^2=k_{b}^{2}$, the kernel of $\widehat{G}$ is not summable and, therefore, the successive approximations produce a diverging series. This implies a non-analytic behavior of
the solution in $\chi_c$. Thus, when a bound state in the radiation continuum
is present, the conventional perturbative approach
becomes inapplicable. Here, a non-perturbative
approach is developed to obtain the solution to the scattering problem
that is valid in any small neighborhood
 of a real pole of the Fourier transform of $\widehat{G}$.

Suppose that the incident radiation is a plane
wave
\begin{equation*}
E_{in}(\rr,t)=2\cos(\kk\cdot \rr-\omega t),\quad
 \kk=\kx\ex+\kz\ez,\quad c k=\omega,
\label{eq:3}
\end{equation*}
where ${\bf e}_{i}$, $i=1,2,3$, denote unit vectors
along the $x$, $y$, and $z$ coordinate axes, respectively.
A general solution to Eq. (\ref{eq:1}) should then
be of the form,
\begin{equation}
E(\rr,t)=\sum_{l=-\infty}^{\infty}  E_{l}(\rr)e^{-i l\omega t}
\label{eq:4}
\end{equation}
where $E_{0}\equiv0$, and for all $l$, $E_{-l}=
\overline{E}_{l}
$ is the complex conjugate of $E_{l}$ (as $E$ is real).
Therefore it is sufficient to determine only
$E_{l},\, l\geq 1$. Next, it is assumed that
the scattering structure is periodic in the $x-$direction
(e.g., a grating). The units of length are chosen so that
the period is one.
Then the amplitudes $E_{l}$ satisfy Bloch's periodicity condition
\begin{equation}
E_{l}(\rr+\ex)=e^{i l\kx}E_{l}(\rr)
\label{eq:5.1}
\end{equation}
This condition follows
from the requirement
that the solution $E$ satisfies the same periodicity
condition as the incident wave $E_{in}$:
\begin{equation*}
E_{in}(\rr+\ex,t)=E_{in}\Bigl(\rr,t-\frac{\kx}{\omega}\Bigr)
\label{eq:5}
\end{equation*}
By Eq.(\ref{eq:1}),
the amplitudes of the different harmonics satisfy the equations,
\begin{equation*}
\Delta E_{l}+l^2k^2\ee E_{l}=-\nu l^2k^2 (\ee-1)\sum_{p} E_{p}E_{l-p},\quad \nu=\frac{\ch}{4\pi(\ec-1)}
\label{eq:6}
\end{equation*}
For ease of notation, the parameter
$\nu$ is often used in lieu of $\ch$. Since
$\nu\sim \ch$, it is a small parameter in the system.

The scattering theory requires that for $l\neq\pm 1$, the partial waves $E_{l}e^{-i l\omega t}$ be outgoing in the spatial infinity ($|z|\rightarrow\infty$). The fundamental waves $E_{\pm 1}e^{\mp i \omega t}$ are a superposition
of an incident plane wave $e^{\pm{i(\kk\cdot\rr-\omega t)}}$ and a scattered wave which is outgoing at the spatial infinity. In all, the above boundary conditions lead to a system of Lippmann-Schwinger integral equations for the amplitudes $E_{l}$:
\begin{equation}
\left\{
\begin{array}{lll}
E_{1} &=&\hv(k^2)[E_{1}+\nu\sum_{p} E_{p}E_{1-p}]+e^{i\kk\cdot \rr}
\\
E_{l}&=&\hv((lk)^2)[E_{l}+\nu\sum_{p}E_{p}E_{l-p}],\quad \,l\geq 2
\end{array}
\right.
\label{eq:7}
\end{equation}
and $E_{-l}=\overline{E}_{l}$  for $l\leq -1$, where
$\hv(q^{2})$ is the integral operator defined by the relation
\begin{equation}
\hv(q^2)[\psi](\rr)=\frac{q^2}{4\pi}\int (\varepsilon(\ro)-1)G_{q}(\rr|\ro)\psi(\ro) d\ro
\label{eq:8}
\end{equation}
in which
$G_{q}(\rr|\ro)$ is the Green's function of the Poisson
operator, $(q^2 +\Delta)G_q(\rr|\ro)=-4\pi \delta(\rr-\ro)$,
with the outgoing wave boundary conditions.
For two spatial dimensions, as in the case considered here
$\rr=(x,z)$ and $\ro=(x_0,z_0)$,
the Green's function is known \cite{mf} to be
$G_{q}(\rr|\ro)=i\pi H_{0}(q|\rr-\ro|)$ where $H_{0}$ is the zero order Hankel function of the first kind.

When $\nu =0$, the amplitudes of all higher harmonics
$(l\geq 2)$ vanish. Therefore it is natural to assume
that $|E_{1}|\gg |E_{2}|\gg |E_{3}|\gg \cdots$
for a small $\nu$. Note that this does not generally
imply that the solution, as a function of $\nu$,
is analytic at $\nu=0$.
Under this assumption, the solution
to the system (\ref{eq:7}) can be approximated
by keeping only the leading terms in each of the series involved. In particular, the first equation in
(\ref{eq:7}) is reduced to
\begin{equation}
E_{1}\approx e^{i\kk\cdot\rr}+\hv(k^2)[E_{1}]+2\nu \hv(k^2)\left[\,\overline{E}_{1}E_{2}\right]
\label{eq:9}
\end{equation}
while the second equation becomes
\begin{equation}
E_{2}\approx \hv((2k)^2)[E_{2}]+\nu \hv((2k)^{2})\left[E_{1}^{2}\right]
\label{eq:10}
\end{equation}
It then follows that a first order approximation to the solution of the nonlinear wave equation (\ref{eq:1}) may be found by solving the system formed by the equations (\ref{eq:9}) and (\ref{eq:10}). To facilitate the subsequent
analysis, the system is rewritten as
\begin{equation}
\begin{cases}
[1-\hv(k^2)][E_{1}]=e^{i\kk\cdot\rr}+2\nu \hv(k^2)\left[\,\overline{E}_{1}E_{2}\right]\\
[1-\hv((2k)^2)][E_{2}]=\nu \hv((2k)^2)\left[E_{1}^{2}\right]
\end{cases}
\label{eq:11}
\end{equation}

Solving the first of Eqs.(\ref{eq:11}) involves inverting the operator $1-\hv(k^2)$, and therefore necessitates a study of the poles of the resolvent $[1-\hv(k^2)]^{-1}$. Such poles are eigenvalues in the generalized eigenvalue problem,
\begin{equation}
\hv(k^{2})[E]=E
\label{eq:12}
\end{equation}
for fixed $\kx$. The corresponding eigenfunctions $E=E_{s}$ are referred to as {\it Siegert states}. In contrast to
Siegert states in quantum scattering theory \cite{b4},
electromagnetic Siegert states satisfy the generalized
eigenvalue problem (\ref{eq:12}) in which the operator
is a nonlinear function of the spectral parameter $k^2$.
Their general properties are studied
in \cite{rs2}.

Eigenvalues have the form $k^2=k_r^2 -i\Gamma$. If $k_r>k_x$, then, according to scattering theory, such a pole is a resonance pole. In the case of the linear
wave equation ($\chi_c = \nu = 0$), the scattered flux peaks at $k = k_r$ indicating the resonance position, whereas
the imaginary part of the pole $\Gamma$  defines the corresponding resonance width (or a spectral width of the scattered flux
peak; a small $\Gamma$  corresponds to a narrow peak).   If  $ \Gamma =0$, the corresponding Siegert state is a {\it bound state}. This is a localized (square integrable) solution of Eq.(\ref{eq:12}). Suppose that the scattering system has a physical parameter $h$ such that a pole $k^2=
k_r^2 (h) - i\Gamma(h)$ of the resolvent
$[1 - \widehat{H}(k^2)]^{-1}$ depends continuously on $h$ and that there is a particular value $h = h_b$ at which the pole becomes
real, i.e., $\Gamma(\hb) = 0$. If $k_{b}=k_r(h_b) > k_x$, then the corresponding bound state lies in the
radiation continuum. Note that Eq.(\ref{eq:12}) may have solutions for real $k^2 < k^2_x$. These are bound states below the
radiation continuum. Such states are not relevant for the present study and, henceforth, bound states are understood
as bound states in the radiation continuum. As noted in the introduction, two periodic planar scattering structures separated by a distance
$2h$ have bound states in the radiation continuum.

Suppose that for fixed $k_x$, the set $k_b^2$ consists
of isolated points (which is generally true)
and the points $(2k_{b})^2$
do not belong to it (which is fulfilled
in a concrete example studied
in the next section \cite{jmp}).
Consequently, the operator $(1-\hv((2k)^2)^{-1}$
is regular in a
neighborhood of $k_{b}^{2}$ and,
for $k$ close to but not equal to $\kb$, the operator $1-\hv(k^2)$ is invertible. It then follows that $E_{1}$ satisfies the nonlinear integral equation,
\begin{equation}
E_{1}=\Bigl(1-\hv(k^2)\Bigr)^{-1}\left[e^{i\kk\cdot\rr}+2\nu^2 \hv(k^2)\left[\overline{E}_{1}
\Bigl(1-\hv((2k)^2)\Bigr)^{-1}\left[\hv((2k)^2)\left[E_{1}^{2}\right]\right]\right]\right]
\label{eq:13}
\end{equation}
where, in accord with the notation introduced in
(\ref{eq:8}), the function on which an operator acts is placed
in the square brackets following the operator.
The operator $\nu^2(1-\hv(k^2))^{-1}$ that determines
the ``nonlinear" part of Eq. (\ref{eq:13})
is not bounded as $k^2\rightarrow k_b^2$, no matter how small
$\nu^2\sim \chi_c^2$ is. This precludes the use of
a power series representation of the solution in $\nu^2$. To find the solution of Eq.(\ref{eq:13}) when $\nu$ is small, its property  under parity transformations is established first.

Suppose that the scattering system is such that
the operator $\hv$ and the parity operator $\widehat{\text{P}}$ defined by $\widehat{\text{P}}[E](x,z)=E(x,-z)$ commute. This implies that the Siegert states have
a specific parity: $\widehat{\text{P}}[E_s]=pE_s$
where $p=\pm 1$.
Consider then the ratio
\begin{equation}
\mu(x,z)=\frac{E_{1}(x,-z)}{E_{1}(x,+z)}=
\frac{\widehat{\text{P}}[E_1]}{E_1}
\label{eq:14}
\end{equation}
It will be proved that $\mu(x,z)\rightarrow p$ in the limit $(h,k)\rightarrow(h_{b},k_{b})$ along a certain curve. Indeed, it follows from the meromorphic expansion of $[1-\hv(k^2)]^{-1}$ that near a pole $k^{2}_{r}(h)-i\Gamma(h)$,
\begin{equation}
E_{1}=\frac{i C(h)}{k^2-k_{r}^{2}(h)+i \Gamma(h)}E_{s}+O(1)
\label{eq:15}
\end{equation}
where $C(h)$ is some constant depending on $h$, and $E_{s}$ is an appropriately normalized Siegert state~\cite{rs2}.
Consider then the curve of resonances $\cb:\, k=k_{r}(h)$
in the $(h,k)$-plane. Along $\cb$,
\begin{equation}
E_{1}(x,z)= \frac{C(h)}{\Gamma(h)} E_{s}(x,z)+O(1)
\label{eq:16}
\end{equation}
Now, as $h\rightarrow\hb$, the width $\Gamma(h)$ goes to $0$, and the Siegert state $E_{s}$ becomes a bound state $E_{b}$ in the radiation continuum.
Equation (\ref{eq:16}) shows that if $C(h)$ does not go to zero faster
than $\Gamma(h)$ as $h\rightarrow \hb$, i.e.,
the pole still gives the leading contribution to $E_1$
in this limit,   then
\begin{equation}
\mu(x,z)\rightarrow\frac{E_{b}(x,-z)}{E_{b}(x,+z)}=
\frac{\widehat{\text{P}}[E_b]}{E_b} = p=
\pm 1
\label{eq:17}
\end{equation}
depending on whether the bound state $E_{b}$
is even or odd in $z$.  For the linear wave equation
($\nu=0$), the constant $C(h)$
is shown to be proportional to $\sqrt{\Gamma(h)}$
\cite{rs2}.
Therefore, for a small $\nu$, the assumption
that $C(h)$ does not go to zero faster than $\Gamma(h)$ as $h\rightarrow \hb$ is justified.

Based on the limit (\ref{eq:17}), the following
(non-perturbative) approach
is adopted to solve
Eq.(\ref{eq:13}) near a bound state.
First, the curve of resonances $\cb$ in the $(h,k)$-plane
is found.
Using the relation (\ref{eq:14}) in the right
side of (\ref{eq:13}), the field $E_1$
is expressed via the ratio $\mu$
with the pair
$(h,k)$ being on the curve $\cb$. Next,
the principal part of the amplitude $E_{1}$
relative to $\Delta h = h-h_b$
is evaluated near a critical point
$(\hb,\kb)$ on $\cb$ by taking $\mu$ to its limit value
(\ref{eq:17}).
This approach reveals a non-analytic dependence
of the amplitude $E_{1}$ on the small parameters $\Delta h$
and $\ch$ of the system and allows to obtain
$E_1$ and $E_2$ when a bound state is present
in the radiation continuum. As the technicalities
of the proposed non-perturbative approach
depend heavily on peculiarities of the scattering
system, the procedure is illustrated with
a specific example.

\section{A periodic double array of subwavelength cylinders}
\label{sec:2}

\begin{figure}[h t]
    \centering
    \includegraphics[scale=0.9]{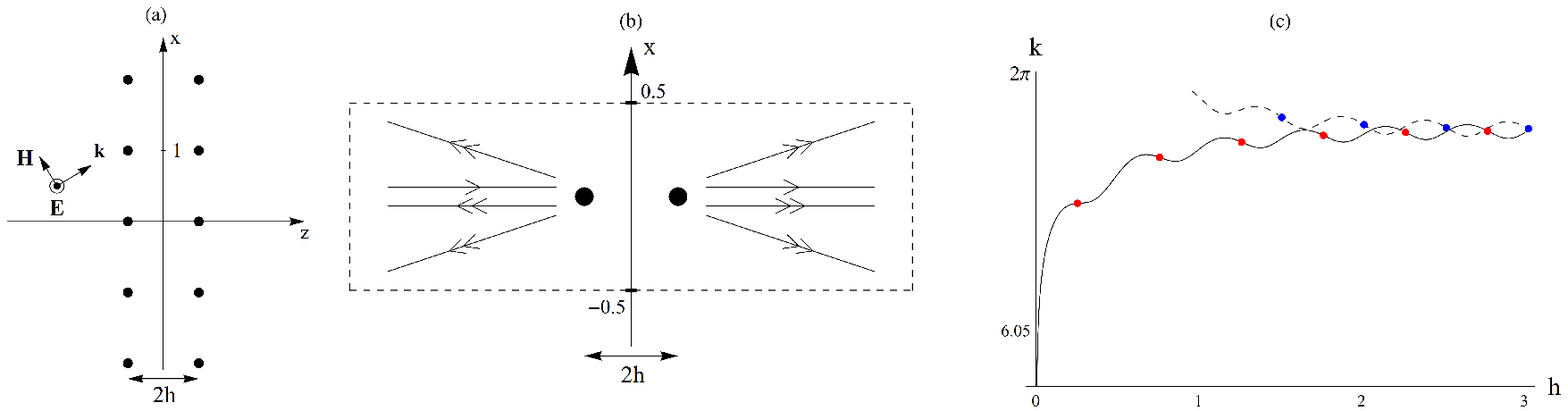}
    \caption{Panel (a): Double array of dielectric cylinders. The unit of length is the array period. The axis of each cylinder is parallel to the $y$-axis, and is at a distance $h$ from the $x$-axis.\newline Panel (b): The scattering process
for  the normal incident radiation ($\kx=0$). The
scattered fundamental harmonic  is symbolized by a single headed arrow while the (generated)
second harmonic radiation is symbolized by a double headed arrow. The incident radiation wave length is such that
only one diffraction channel is open for
the fundamental harmonic while
three diffraction channels are open for the second harmonic.
 The flux measured through the faces $L_{\pm 1/2}: x=\pm 1/2$ cancels out due to the Bloch periodicity condition
as explained in Appendix \ref{asec:2}. \newline Panel (c): The solid and dashed curves show the position
(frequency $\omega_r=ck_r$)
of scattering resonances as functions of the distance between
the arrays, $k=k_{r}(h)$. The dots on the curves
indicate positions of bound states in the radiation continuum (i.e., the values
of $h$ at which a resonance turns into a bound state).
The solid curve connects bound states symmetric
relative to the reflection $z\rightarrow-z$. The dashed line connects the skew symmetric bound states. The curves are realized for $R=0.08,\, \ec=2$, and $\kx=0$ (normal incidence).}
    \label{fig:1}
    \end{figure}

The system considered is sketched in Fig.~\ref{fig:1}(a). It consists of an infinite double array of parallel, periodically positioned cylinders. The cylinders are made of a nonlinear dielectric material with a linear dielectric constant $\ec>1$, and a second
order susceptibility $\ch\ll 1$. The coordinate system is set so that the cylinders are parallel to the y-axis, the structure is periodic along the x-axis, and the z-axis is normal to the structure. The unit of length is taken to be the array period, and the distance between the two arrays relative to the period is $2h$.

The solution
of the integral equation (\ref{eq:13}) is obtained
for $k^2$ near $k_b^2$ in the limit of subwavelength dielectric cylinders. The approximation is defined by a small parameter
\begin{equation}
\dl(q)=\frac{(q R)^2}{4}(\ec-1)\ll 1
\label{eq:18}
\end{equation}
which
is the scattering phase of a plane wave with the wavenumber
$q$ on a single cylinder of radius $R$.
For sufficiently small $R$, this approximation is
justified. The integral kernel of $\hv(q^2)$ is defined
by (\ref{eq:8}) and has support
on the region occupied by cylinders. The
condition (\ref{eq:18}) implies that the wavelength
is much larger than the radius $R$, and therefore
field variations within each cylinder may be neglected, so that
$\psi (x,z)\approx \psi(n,\pm h)$ where
$(n,\pm h)$ are the positions of the axes of the cylinders ($n$ is an integer).
The integration in $\hv(q^2)[\psi]$
yields then an infinite sum
over positions of the cylinders.
By Bloch's condition, $\psi(n,\pm h) =
e^{inq_x}\psi(0,\pm h)$, so that the
function $\hv(q^2) [\psi](x,z)$ is fully determined
by the two values $\psi(0,\pm h)$.
In particular,
\begin{equation}
\hv(q^2)[\psi](0,\pm h)
\approx \alpha\psi(0,\pm h)+\beta\psi(0,\mp h)
\label{eq:19}
\end{equation}
where the coefficients $\alpha$ and $\beta$
are shown to be \cite{jmp}
\begin{subequations}\label{eq:20}
\begin{equation*}\label{eq:20.1}
\alpha(q,q_{x})=2\pi i \dl(q)\left( \sum_{m=-\infty}^{\infty} \left(\frac{1}{q_{z,m}}-\frac{1}{2\pi i(|m|+1)}\right) +\frac{i}{\pi}\ln (2\pi R)\right)
\end{equation*}
\begin{equation*}\label{eq:20.2}
\beta(q,q_{x},h)=2\pi i \dl(q) \sum_{m=-\infty}^{\infty} \frac{e^{2i h q_{z,m}}}{q_{z,m}}
\end{equation*}
\end{subequations}
where $q_{z,m}=\sqrt{q^2-(q_{x}+2\pi m)^2}$ with the convention that if $q^2<(q_{x}+2\pi m)^2$, then $q_{z,m}=i\sqrt{(q_{x}+2\pi m)^2-q^2}$.
To obtain
the energy flux scattered by the structure, the action of the operator $\hv(q^2)$ on $\psi$  must be determined in the asymptotic
region $|z|\rightarrow\infty$.
It is found that for $|z|>h+R$,
\begin{equation}
\hv(q^2)[\psi](x,z)\approx 2\pi i\dl(q) \sum_{m=-\infty}^{\infty}\left(\psi(0,h)e^{i|z-h|q_{z,m}}+\psi(0,-h)
e^{i|z+h|q_{z,m}}\right) \frac{e^{i x(q_{x}+2\pi m)}}{q_{z,m}}
\label{eq:21}
\end{equation}

\section{Amplitudes of the fundamental and second harmonics}
\label{sec:3}

Now that the action of the operator
$\hv(q^2)$ has been established in (\ref{eq:19}) and
(\ref{eq:21}), the amplitudes $E_{1}$ and $E_{2}$ of the fundamental and second harmonics can be determined by solving the system
(\ref{eq:11}). As noted earlier, this will be done along a curve $\cb$ in the $h,k$-plane defined by $k=k_{r}(h)$ where $k_{r}(h)$ is the real part of a pole of $[1-\hv(k^2)]^{-1}$, or equivalently, when the incident radiation
has the resonant wave number $k=k_{r}(h)$.
To find the curve, the eigenvalue problem
 (\ref{eq:12}) is solved in
in the approximation (\ref{eq:19}):
\begin{equation}
  [1-\hc]\begin{pmatrix}
         E_{b+}\\
         E_{b-}
        \end{pmatrix}=\begin{pmatrix}
                        0\\
                        0
                      \end{pmatrix},\quad \hc=\begin{pmatrix}
                                             \al&\be\\
                                             \be&\al
                                             \end{pmatrix}
\label{eq:22}
\end{equation}
where $E_{b\pm}=E_{b}(0,\pm h)$ and
the functions $\al=\al(k,\kx)$ and $\be=\be(k,\kx)$ have been defined in the previous section. In particular, bound states occur at the points $(\hb,\kb)$ at which the determinant
 $\det(1-\hc)$ vanishes,
\begin{equation*}
\det\begin{pmatrix}
1-\al&-\be\\
-\be &1-\al
\end{pmatrix}
=(1-\al-\be)(1-\al+\be)=0
\label{eq:23}
\end{equation*}
It follows from Eq.(\ref{eq:22}) that the bound states
for which $1-\al-\be=0$ are even in $z$ because
$E_{b+}=E_{b-}$ in this case. Similarly, the bound states for
 which $1-\al+\be=0$ are odd in $z$. More generally, the poles of the resolvent $[1-\hv(k^2)]^{-1}$ are
complex zeros of $\det(1-\hc)$. They are found by the
conventional scattering theory formalism. Specifically, the resonant wave numbers $k^2=k^2_{r}(h)$ are obtained
by solving the equation $\re\{1-\al\pm\be\}=0$ for
the spectral parameter $k^2$.
According to the convention adopted in the representation
(\ref{eq:15}),
the corresponding resonance widths are defined by
\begin{equation*}
\Gamma(h)=-\frac{\im\{1-\al\pm\be\}}{\partial_{k^2} \re\{1-\al\pm\be\}}\bigg|_{k^2=k^2_{r}(h)}
\label{eq:24}
\end{equation*}
where $\partial_{k^2}$ denotes the derivative with respect to
$k^2$. This definition of the width corresponds to the linearization of $\re\{1-\al\pm\be\}$ near $k^2=k^2_r(h)$
as a function of $k^2$ in the pole factor
$[1-\al\pm\be]^{-1}$.
The curves of resonances $k=k_{r}(h)>\kx$ come in pairs. There is a curve connecting the symmetric bound states in the $h,k$-plane, and
another curve that connects the odd ones.

In what follows, only
the curve connecting symmetric bound states will be
considered. The other curve can be treated similarly.
Panel (c) of Fig.~\ref{fig:1} shows that the first
symmetric bound state occurs when the distance $2h$
is about half the array period, while the skew-symmetric
bound states emerge only at larger distances.
This feature is explained in detail in \cite{jmp}.
So, the solution obtained near the first symmetric
bound state corresponds to the smallest possible transverse
dimension of the system (roughly a half of the wave
length of the incident radiation).
Thus, from now on the curve of resonances $\cb$ refers to the curve in the $h,k$-plane defined
by the equation $\re\{1-\al-\be\}=0$.
To simplify the technicalities,
it will be further assumed that only one diffraction channel is open for the fundamental harmonics,
i.e., $\kx<k<2\pi-\kx$.

Let $k=k_{r}(h)$ be the solution of
$\re\{1-\al-\be\}=0$. By making use of the explicit
form of the functions $\alpha$ and $\beta$
for one open diffraction channel, one infers that
along the curve $k=k_{r}(h)$,
\begin{equation*}
1-\al-\be=i \im\{1-\al-\be\}=-i\frac{4\pi \dl(k_b)}
{k_{z}}\vp^2,\,\quad \vp=\cos(h k_{z}),\ \quad
k_{z}=\sqrt{k^2_b -k_x^2}
\label{eq:25}
\end{equation*}
Bound states in the radiation continuum occur
when the distance $h$ satisfies the equation $\vp=0$, i.e.,
$h\sqrt{k_{r}^2(h)-k_x^2}=(n-1/2)\pi$ with $n$ being a
positive integer. Its solutions $h=\hb(n)$ define
the corresponding values of the wave numbers of the bound states, $k_b(n)=k_{r}(\hb(n))$. So, the sequence of pairs
$\{(\hb(n),\kb(n))\}_{n=1}^{\infty}$ indicates positions
of the bound states on the curve $\cb$.
In the limit $h\rightarrow \hb(n)$ along $\cb$,
the function $\vp$ has the asymptotic behavior,
\begin{equation*}
\vp = (-1)^{n}k_{z,b}\Delta h + o(\Delta h),\quad
k_{z,b}=\kz|_{h=h_{b}(n)},\quad \Delta h=h-\hb
\label{eq:26}
\end{equation*}
The objective is to determine the dependence of the amplitudes $E_{1}$ and $E_{2}$ on the parameters $\Delta h$ and $\ch$ which are both small.

To this end, let $E_{1\pm}=E_{1}(0,\pm h)$ be the values of the field $E_{1}$ on the axes of the cylinders
at $(0,\pm h)$, and $E_{2\pm}=E_{2}(0,\pm h)$ be the values of the field $E_{2}$ on the same cylinders.
In the subwavelength approximation, these values
determine the scattered field because the latter
is produced by the radiation of point dipoles induced
by the incident wave on the scatterers
and the strength of the dipoles
is proportional to $E_{1\pm}$ for the fundamental
harmonic and $E_{2\pm}$ for the second harmonic.
Applying the rule (\ref{eq:19}) to evaluate the action
of the operator $\hv(q^2)$ in the system (\ref{eq:11}),
the first equation of the latter becomes,
\begin{subequations}\label{eq:27}
 \begin{equation}\label{eq:27.1}
  [1-\hc]
  \begin{pmatrix}
  E_{1+}\\
  E_{1-}
  \end{pmatrix}
  =2\nu\hc
        \begin{pmatrix}
        \overline{E}_{1+}E_{2+}\\
        \overline{E}_{1-}E_{2-}
        \end{pmatrix}+\begin{pmatrix}
                      e^{+i h\kz}\\
                      e^{-i h\kz}
                      \end{pmatrix}
    \end{equation}
Similarly, the second of Eqs.(\ref{eq:11}) yields,
\begin{equation}\label{eq:27.2}
  [1-\hcd]
  \begin{pmatrix}
  E_{2+}\\
  E_{2-}
  \end{pmatrix}
  =\nu \hcd
      \begin{pmatrix}
      E_{1+}^{2}\\
      E_{1-}^{2}
      \end{pmatrix},\quad \hcd=\hc(2k,2\kx)
  \end{equation}
  \end{subequations}
As stated above, the resolvent $[1-\hv((2k)^2)]^{-1}$ is regular in a neighborhood of $\kb$ so that
Eq. (\ref{eq:27.2}) can be solved for $E_{2\pm}$,
which defines the latter as functions of
$E_{1\pm}$. The substitution of
this solution into Eq.(\ref{eq:27.1}) gives a system
of two nonlinear equations
for the fields $E_{1\pm}$. Adding these
equations and replacing the field $E_{1-}$
by its expression $E_{1-}=\mu(0,h)E_{1+}$ in terms of the field ratio of Eq.(\ref{eq:14}) yields the following implicit relation between the field $E_{1+}$ and
its amplitude $|E_{1+}|$:
\begin{equation}
E_{1+}=-\frac{\vp}{\frac{\nu^2}{\zeta}|E_{1+}|^2+\vp^2\xi}
\label{eq:28}
\end{equation}
where $\vp=\cos(h\kz)$ and $\nu=\frac{\ch}{4\pi(\ec-1)}$ are small and, in terms of the field ratio $\mu\equiv
\mu(0,h)$, the values of $\zeta$ and $\xi$ read,
\begin{equation}
\xi=i\frac{2\pi \dl(k)}{\kz}(1+\mu),\quad
\frac{1}{\zeta}=\left(1+i\frac{4\pi\dl(k)}{\kz}\vp^2\right)\left(a+b\mu^2+\overline{\mu}\left(b+a\mu^2\right)\right)
\label{eq:29}
\end{equation}
with $a$ and $b$ being defined by the relation,
\begin{equation}
[1-\hcd]^{-1}\hcd=\begin{pmatrix}
                    a&b\\
                    b&a
                    \end{pmatrix}
                    \label{eq:29.1}
                    \end{equation}
In particular,  $\zeta$ and $\xi$ are
continuous functions of $\mu$ and $\vp$.
In Appendix~\ref{asec:1} it is shown that if $\zeta_{b}$ and $\xi_{b}$ are the respective limits of $\zeta$
and $\xi$ as $h\rightarrow \hb$ along the curve of resonances $\cb$, then these limits are nonzero. It follows then
that Eq.(\ref{eq:28}) for $E_{1+}$ is singular in both $\nu$ and $\vp$ when these parameters are small, i.e., in the limit $(\nu,\vp)\rightarrow (0,0)$. Furthermore, there is no way to solve the said equation perturbatively in either of the parameters. A full non-perturbative solution can be obtained using Cardano's method for solving cubic polynomials. Indeed, by taking the modulus squared of both sides of the equation, it is found that,
\begin{equation}
X^{3}+2\frac{\vp^2}{\nu^2}\re\{\zeta\xi\}X^2+\frac{\vp^4}{\nu^4}|\zeta\xi|^{2}X-\frac{\vp^2}{\nu^4}|\zeta|^2=0,\quad
X=|E_{1+}|^2
\label{eq:30}
\end{equation}

The solution to this cubic equation is obtained in Appendix~\ref{asec:3}. It is proved there that Eq. (\ref{eq:30})
admits a unique real solution so that there is no ambiguity on the choice of $E_{1+}$. In the vicinity of a point $(\hb,\kb)$ along the resonance curve $\cb$, the field $E_{1+}$ is found to behave as,
\begin{equation}
|E_{1+}|=\frac{|\Delta h|^{1/3}}{\ch}\,\tau(\Delta h,\ch)
\label{eq:31}
\end{equation}
Recall that $\Delta h=h-\hb$.
An explicit form of the
function $\tau(\Delta h,\ch)$ is given in Appendix
\ref{asec:3} (see Eq. (\ref{eq:b8})).
It
involves combinations of the square and cube roots of functions in $\Delta h$ and $\ch$ and has the property
that $\tau(\Delta h,\ch)\rightarrow 0$ as $(\Delta h,\ch)\rightarrow (0,0)$ (in the sense of the two-dimensional
limit). In the limit $h\rightarrow h_b$, the field
ratio $\mu$ approaches $1$ for a symmetric bound state
as argued earlier. Therefore it follows from
Eqs.(\ref{eq:27.2}) that $E_{2\pm}\sim \nu E_{1+}^2$
because the matrix (\ref{eq:29.1}) exists at $h=h_b$.
Since $\nu\sim \chi_c$, relation (\ref{eq:31}) leads
to the conclusion that
\begin{equation*}
\frac{E_{2\pm}}{E_{1+}}=O(|\Delta h|^{1/3})
\label{eq:32}
\end{equation*}
Thus, the approximation $|E_{1}|\gg |E_{2}|$
used to truncate the system (\ref{eq:11})
remains valid for $h$ close to the critical value $\hb$
despite the non-analyticity of the amplitudes
 at $(\Delta h,\chi_c)=(0,0)$.

\section{Flux analysis: the conversion efficiency}
\label{sec:4}

For the nonlinear system considered, even though Poynting's theorem takes a slightly different form as compared to linear Maxwell's equations, the flux conservation for the time averaged Poynting vector holds. The scattered energy flux
carried across a closed surface by each of the different harmonics adds up to the incident flux across that surface. The flux conservation theorem is stated
in Appendix \ref{asec:2}. Consider a closed surface
that consists of four faces,
$L_{\pm}=\{(x,z)|-\frac{1}{2}\leq x\leq \frac{1}{2}, z\rightarrow\pm\infty\}$ and $L_{\pm 1/2}=
\{(x,z)| x=\pm 1/2\}$ as depicted in Fig.~\ref{fig:1}(b).
As argued in Appendix \ref{asec:2}, the scattered
flux of each $l^{\text{th}}$-harmonics across the union
of the faces
$L_{\pm 1/2}$ vanishes because of the Bloch condition
(and so does the incident flux for any $k_x$).
Therefore only the flux conservation across the union
of the faces $L_\pm$ has to be analyzed.
If $\sigma_{l}$ designates the ratio of the
scattered flux carried
 by the $l^{\text{th}}$-harmonics across the faces $L_\pm$
to the incident flux across the same faces,
then $\sum_{l\geq 1}\sigma_{l}=1$. Thus, for $l\geq 2$,
$\sigma_{l}$ defines the {\it conversion ratio} of fundamental harmonics into the $l^{\text{th}}$-harmonics.

In the perturbation theory used here, only the ratios $\sigma_{1}$ and $\sigma_{2}$ may be evaluated. By laborious calculations it can be shown that $\sigma_{1}+\sigma_{2}\leq 1$ as one would expect (see Appendix~\ref{asec:2} for details). Hence, the efficiency of converting the fundamental harmonic into the second harmonic is simply determined by
the maximum value of $\sigma_{2}$ as a function of
the parameter $h$ at a given value of the nonlinear susceptibility $\chi_c$.

The ratio $\sigma_2$ is defined in terms of the
scattering amplitudes
of the second harmonic, i.e., by the amplitude
of $E_2$ in the asymptotic region $|z|\rightarrow\infty$:
\begin{equation}
E_{2}(\rr)\rightarrow
           \begin{cases}
           \displaystyle{\sum_{m^{op,sh}} R_{m}^{sh} e^{i \rr\cdot\mathbf{k}_{m,sh}^{-}}},\quad z\rightarrow -\infty\\
           \displaystyle{\sum_{m^{op,sh}} T_{m}^{sh} e^{i \rr\cdot \mathbf{k}_{m,sh}^{+}}},\quad z\rightarrow +\infty
           \end{cases}
           \label{eq:33}
           \end{equation}
where $\mathbf{k}_{m,sh}^{\pm}=
(2k_{x}+2\pi m )\ex\pm k_{z,m}^{sh}\ez$ is the wave vector
of the second harmonic in the
$m^{\text{th}}$ open diffraction channel. Recall
that the $m^{\text{th}}$
channel is open provided
 $(2k)^2>(2\kx+2\pi m)^2$ and in this case
$k_{z,m}^{sh}=\sqrt{(2k)^2-(2\kx+2\pi m)^2}$, while if the channel is closed, then $k_{z,m}^{sh}=i \sqrt{(2\kx+2\pi m)^2-(2k)^2}$. In the asymptotic region $|z|\rightarrow\infty$, the field in closed channels decays exponentially and,
hence, the energy flux can only be carried in
open channels to the spatial infinity.
The summation in Eqs.(\ref{eq:33})
is taken only over those values of $m$ for which
the corresponding diffraction channel is open for
the second harmonic, which
is indicated by
the superscript ``$op,sh$'' in the summation index
$m^{op,sh}$.  Note that there is more than one open diffraction channel for the second harmonic even though only one diffraction channel is open for
the fundamental one. For instance, if the $x-$component
of the wave vector $\kk$, i.e., $\kx$,
is less than $\frac{\pi}{2}$, there are 3 open diffraction channels for the second harmonic,
the channels $m=0,\, m=-1$, and $m=1$.
These three directions of the wave vector of the second
harmonic propagating
in each of the asymptotic regions $z\rightarrow\pm\infty$
are depicted in Fig.~\ref{fig:1}(b) by double-arrow rays.
Thus,
in terms of the scattering amplitudes introduced
in Eqs.~(\ref{eq:33}), the ratio of the second harmonic
flux across $L_\pm$ to the incident flux
is
\begin{equation*}
\sigma_{2}=\frac{1}{2\kz}\sum_{m^{op,sh}} k_{z,m}^{sh}\left(|R_{m}^{sh}|^2+|T_{m}^{sh}|^2\right)
\label{eq:34}
\end{equation*}

The scattering amplitudes $R_{m}^{sh}$ and $T_{m}^{sh}$ are inferred from Eq.(\ref{eq:10}) in which
the rule (\ref{eq:21}) is applied
to calculate the action of the operator $\hv((2k)^2)$
in the far-field regions $|z|\rightarrow\infty$:
\begin{equation*}
\begin{cases}
\displaystyle{R_{m}^{sh}=\frac{2\pi i\dl(2k)}{k_{z,m}^{sh}}\left[(E_{2+}+\nu E_{1+}^{2}) e^{i h k_{z,m}^{sh}}+(E_{2-}+\nu E_{1-}^{2}) e^{-i h k_{z,m}^{sh}}\right]}\vspace{0.1 cm}\\
\displaystyle{T_{m}^{sh}=\frac{2\pi i\dl(2k)}{k_{z,m}^{sh}}\left[(E_{2+}+\nu E_{1+}^{2}) e^{-i h k_{z,m}^{sh}}+(E_{2-}+\nu E_{1-}^{2}) e^{i h k_{z,m}^{sh}}\right]}
\end{cases}
\label{eq:35}
\end{equation*}

Since for $\nu\neq0$, the amplitudes $E_{1\pm}$ remain finite as $h\rightarrow \hb$ along $\cb$, and since $\mu\rightarrow 1$ in the said limit; the principal part of $\sigma_{2}$
in a vicinity of a bound state along the curve $\cb$
is obtained by setting $E_{1-}=E_{1+}$ in Eq.
(\ref{eq:27.2}), solving the latter for $E_{2\pm}$,
and substituting the solution into the expression
for $\sigma_2$. The result reads
\begin{equation}
\sigma_{2}=C_{b} \nu^{2}|E_{1+}|^4
\label{eq:36}
\end{equation}
where $C_{b}$ is a constant obtained by taking all nonsingular factors in the expression of $\sigma_{2}$ to their limit as $h\rightarrow \hb$, which gives
\begin{equation*}
C_{b}=\left[\frac{(16\pi\dl(k))^{2}}{\kz}\left|1+a+b\right|^{2}\sum_{m^{op,sh}} \frac{\cos^{2}(h k_{z,m}^{sh})}{k_{z,m}^{sh}}\right]_{(h,k)=(\hb,\kb)}
\label{eq:37}
\end{equation*}
for $a$ and $b$ defined in Eq.(\ref{eq:29.1}).
Using the identity $|E_{1+}|^4 = |E_{1+}|^2|E_{1+}|^2$, and substituting Eq.(\ref{eq:28}) into one of the factors
$|E_{1+}|^2$, the conversion ratio $\sigma_{2}$ is
expressed as a function of a single
real variable,
\begin{equation}
\sigma_{2}(u)=C_{b}^{\prime} \frac{u}{|u+\zeta_{b}\xi_{b}|^{2}},\, \quad u=\left(\frac{\nu |E_{1+}|}{\vp}\right)^{2}
\label{eq:38}
\end{equation}
where $C_{b}^{\prime}=C_{b}|\zeta_{b}|^2$ is a constant, and $\xi$ and $\zeta$ in Eq.(\ref{eq:29}) have been taken at their limits as $h\rightarrow \hb$ to obtain the principal part of $\sigma_{2}$. The function $u\mapsto \sigma_{2}(u)$ on $[0,\infty)$ is  found to attain its absolute maximum
at $u=|\xi_{b}\zeta_{b}|$. This condition determines
the distance $2h$ between the arrays at which
the conversion rate is maximal
for given parameters $R,\, \ec$ and $\ch$ of the system. Indeed, since $u\sim |E_{1+}|^2$ should also
satisfy the cubic equation (\ref{eq:30}),
the substitution of $u=|\zeta_{b}\xi_{b}|$ into the latter
yields the condition
\begin{subequations}\label{eq:39}
\begin{equation}\label{eq:39.1}
\frac{\nu^2}{\vp^4}=2|\xi_{b}|^2\left(|\xi_{b}\zeta_{b}|+\re\{\xi_{b}\zeta_{b}\}\right)
\end{equation}
In particular, in the leading order in $\dl(k)$, the optimal distance $2h$ between the two arrays is given by the formula,
\begin{equation}\label{eq:40}
(h-\hb)^4=\frac{\chi_{c}^{2}}{8\pi^5 k_{z,b}(\kb R)^6(\ec-1)^5}
\end{equation}
\end{subequations}
where as previously, $k_{z,b}=\sqrt{k_{b}^{2}-k_{x}^{2}}$.

\begin{figure}[h t]
    \centering
    \includegraphics[scale=0.9]{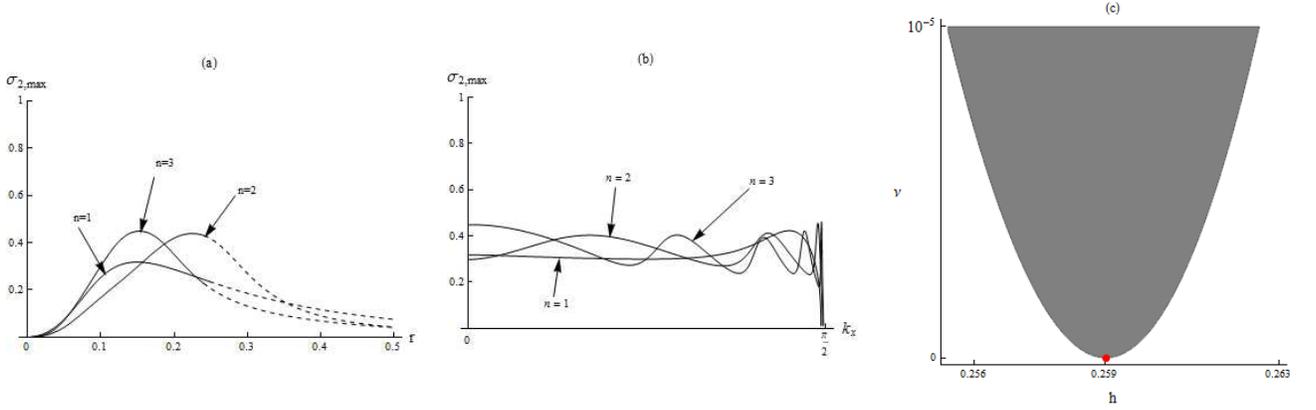}
    \caption{Panel (a): The conversion efficiency is plotted against the cylinder radius $R$ for the critical points $(\hb(n),\kb(n)),\, n=1,2,3$, and $\ec=1.5$ and $\kx=0$
(the normal incidence).
The dashed parts of the curves
indicate the regions where $\dl(k)> 0.25$ and, hence,
$\dl(2k)>1$, i.e., the subwavelength approximation
becomes inapplicable for the second harmonic.
\newline Panel (b): The conversion efficiency is plotted against $\kx$ for the critical points $(\hb(n),\kb(n)),\, n=1,2,3$. The curves are realized for $R=0.15$ and $\ec=1.5$.\newline Panel (c): The region of validity of
the developed theory for the first bound state
$\hb(1)\approx 0.259$. The shadowed part of the
$(\nu, h)-$plane is defined by the condition
$\tau_++\tau_- >0$ under which,
according to Eq.(\ref{eq:b8}), the amplitude
$|E_{1+}|$ exists and unique as explained in
 Appendix \ref{asec:3}.
The plot is realized for $R=0.1,\, \ec=2$ and $\kx=0$.
The parabola-like curve is an actual boundary of the
shadowed region; the top horizontal line represents
no restriction. So there is a wide range of the physical
and geometrical parameters within the shadowed
region which satisfy (\ref{eq:39.1}).
The regions of validity for other bound states looks
similar.}
    \label{fig:2}
    \end{figure}

The maximum value $\sigma_{2,{\rm max}}$ of the conversion ratio $\sigma_{2}$ is the sought-for {\it conversion efficiency}. An interesting feature to note is that
$\sigma_{2,{\rm max}}=\sigma_2(|\zeta_{b}\xi_{b}|)$
is independent of the nonlinear susceptibility $\chi_{c}$
because the constants $C_{b}^{\prime}$, $\xi_b$, and $\zeta_b$
are fully determined by the position of the bound state
$(k_b, h_b)$.
In other words, if the distance
between the arrays is chosen to satisfy the condition
(\ref{eq:39.1}), the conversion efficiency
$\sigma_{2,{\rm max}}$ is the same for a wide range
of values of the nonlinear susceptibility $\chi_{c}$.
This conclusion
follows from two assumptions made in the analysis.
First, the subwavelength approximation should be valid for both the fundamental and
second harmonics, i.e., the radius of cylinders should be
small enough. Second, the values of $h-h_b$ and $\nu$
(or $\chi_c$) must be such that the analysis of
the existence and uniqueness of $|E_{1+}|$ given in
Appendix \ref{asec:3} holds, that is, Eq. (\ref{eq:30})
should have a unique real solution under the
condition (\ref{eq:39.1}). The geometrical and physical
parameters of the studied system can always be chosen
to justify these two assumptions as illustrated in
Fig.~\ref{fig:2}.

Panels (a) and (b) of
Fig. \ref{fig:2} show the conversion efficiency
$\sigma_{2,{\rm max}}$ for
the first three symmetric bound states $n=1,2,3$,
as, respectively, a function of the cylinder radius $R$
when $k_x=0$ and of $k_x$ when $R=0.15$.
For all curves presented in the panels,
$\varepsilon_c=1.5$.
The values of $\sigma_{2,{\rm max}}$ are
evaluated numerically by Eq. (\ref{eq:38})
where $u=|\xi_b\zeta_b|$.
The solid parts of the curves in Panel (a) correspond to
the scattering phase $\delta_0(k)<0.25$ with
$k= k_b$. Note that the wavelength at which
the second harmonic generation is most efficient
is the resonant wavelength defined by $k=k_r(h)$ where
$h$ satisfies the condition (\ref{eq:39.1}).
For a small $\chi_c$, the scattering phase
at the resonant wavelength can well be approximated
as $\delta_0(k)\approx \delta_0(k_b)$. The condition
$\delta_0(k_b)<0.25$ ensures that the scattering
phase for the second harmonic satisfies the
inequality $\delta_0(2k_b)=4\delta_0(k_b)<1$
otherwise the validity of the subwavelength approximation
cannot be justified. The dashed parts of the curves
in Panels (a) of Fig.~\ref{fig:2} correspond
to the region where $\delta_0(k_b)> 0.25$. Panels (a) and (b)
of the figure
show that the conversion efficiency can be as
high as 40\% for a wide range of the incident angles
and values of the cylinder radius.
Such a conversion efficiency is comparable with that
 achieved in optically nonlinear crystals
at a typical beam propagation length (active length)
of a few centimeters,
whereas here the transverse dimension $2h$ of the system
studied here can be as low as a half of the wavelength, i.e., for
an infrared incident radiation, $2h$ is about a few
hundred nanometers. Indeed, as one can see in
Fig.~\ref{fig:1}(c), the first
bound state occurs at $h_b(1)\approx 0.259$ and $k_b
\approx 2\pi$ which corresponds to the wavelength
$\lambda_b = 2\pi/k_b\approx 1$.

The stated conversion efficiency can be fairly well estimated
in the leading order of $\dl(k)$:
\begin{equation}
\sigma_{2,{\rm max}}=\sigma_{2}(|\zeta_{b}\xi_{b}|)\approx 8\pi\dl(k)\sum_{m^{op,sh}} \frac{\cos^{2}(h k_{z,m}^{sh})}{k_{z,m}^{sh}}\Bigg|_{(h,k)=(\hb,\kb)}
\label{eq:41}
\end{equation}
Suppose that only one diffraction channel
is open for the incident radiation. Then $m=0,\pm 1$
in Eq.(\ref{eq:41}) (three open channels for the
second harmonics). Let $\sigma_{2,{\rm max}}^{0}$
denote the term $m=0$, i.e.
$\sigma_{2,{\rm max}}^{0}$ is the second harmonic flux
in which the contribution of the
channels with
$m=\pm 1$ is omitted. In particular, $\sigma_{2,{\rm max}}^{0}<
\sigma_{2,{\rm max}}$.  One infers from Eq. (\ref{eq:41}) that,
\begin{equation*}
\sigma_{2,{\rm max}}^{0}\approx \frac{4\pi\dl(\kb)}{k_{z,b}}\cos^{2}(2 \hb k_{z,b})
\label{eq:42}
\end{equation*}
As the pair $(\hb,\kb)$ at which a symmetric bound state is formed satisfies the equation $\cos(\hb k_{z,b})=0$, it follows that $\cos(2\hb k_{z,b})= -1$. Hence,
\begin{equation*}
\sigma_{2,{\rm max}}^{0}\approx \frac{k_{b}^{2}}{k_{z,b}} \pi R^2(\ec-1)
\label{eq:43}
\end{equation*}
The wavenumbers $\kb$ at which the bound states occur lie just below the diffraction threshold $2\pi-\kx$, i.e., $\kb\lessapprox 2\pi-\kx$ (see Fig.~\ref{fig:1}(c)
and \cite{jmp} for details).
So that in the case of normal incidence ($\kx=0$), the above estimate  becomes,
\begin{equation*}
\sigma_{2,{\rm max}}^{0}\approx 2\pi (\pi R^2)(\ec-1)
\end{equation*}
with $\delta_{0}(2\pi)\ll 1$. If, for instance, $R=0.15$ and $\ec=2$, then,
\begin{equation*}
\sigma_{2,{\rm max}}^{0}\approx 44\%
\label{eq:45}
\end{equation*}
for $\delta_{0}(2\pi)\approx 0.22$.

\appendix
\begin{center}
{\Large\bf Appendices}
\end{center}

\section{Estimation of $\zeta$ and $\xi$}
\label{asec:1}

Here the limit values $\zeta_{b}$ and $\xi_{b}$ of the functions $\zeta$ and $\xi$ defined in Eq. (\ref{eq:29})
are estimated as $h\rightarrow h_b$
along the resonance curve $\cb$.
For $\xi_{b}$ this is immediate. Indeed, in the aforementioned limit, the field ratio
$\mu\rightarrow 1$ for a symmetric bound state and, therefore,
\begin{equation*}
\xi_{b}=i\frac{4\pi \dl(k)}{\kz}\Bigg|_{k=\kb}
\label{eq:a1}
\end{equation*}
For $\zeta_{b}$, the estimate follows from
that wave numbers $\kb$ at which bound states exist are close to the diffraction threshold $2\pi-\kx$
when only one diffraction channel is open
for the fundamental harmonic, i.e., $\kx<k<2\pi-\kx$
\cite{jmp}.
 Indeed, in the first order of $\dl(k)$,
\begin{equation}
k_{b}\approx 2\pi-\kx-\frac{8\pi^2\delta_{0}^{2}(2\pi-\kx)}{2\pi-\kx}
\label{eq:a2}
\end{equation}
This proximity of the wavenumbers $\kb$ to the diffraction threshold $2\pi-\kx$ allows one to determine the leading terms in the coefficients $a$ and $b$ defined in Eq.(\ref{eq:29.1}), and hence $\zeta_{b}$. To proceed, the coefficients $\ald=\alpha(2k,2\kx)$ and $\bed=\beta(2k,2\kx,h)$ of the matrix $\hcd$ are rewritten by separating
explicitly the real and imaginary parts:
\begin{subequations}\label{eq:a3}
\begin{equation}\label{eq:a3.1}
\ald+\bed=\psi_{+}+i S_{c},\quad \ald-\bed=\psi_{-}+i S_{s}
\end{equation}
where $S_{c}$ and $S_{s}$ are defined by the relations,
\begin{equation}\label{eq:a3.2}
S_{c}=16\pi\dl(k)\sum_{m^{op,sh}} \frac{\cos^{2}(h k_{z,m}^{sh})}{k_{z,m}^{sh}},\quad S_{s}=16\pi\dl(k)\sum_{m^{op,sh}} \frac{\sin^{2}(h k_{z,m}^{sh})}{k_{z,m}^{sh}}
\end{equation}
\end{subequations}
and the index $m^{op,sh}$ indicates that the summations are to be taken over all open diffraction channels for the second harmonic. Using the estimate (\ref{eq:a2}), the functions $\psi_{\pm}$ are found to obey the estimates,
\begin{equation*}
\psi_{+}=2+O(\delta_{0}(\kb)),\quad\, \psi_{-}=O(\delta_{0}(\kb))
\label{eq:a5}
\end{equation*}
These expressions are then used to estimate $\zeta_{b}^{-1}=2(a+b)$. In the first order of $\delta_{0}(\kb)$ one infers
that
\begin{equation}
\zeta_{b}\approx -\frac{1}{4}-2\pi i \dl(k)\sum_{m^{op,sh}} \frac{\cos^{2}(h k_{z,m}^{sh})}{k_{z,m}^{sh}}\Bigg|_{(h,k)=(\hb,\kb)}
\label{eq:a6}
\end{equation}

\section{Complements on the flux analysis: Flux conservation}
\label{asec:2}

For the nonlinear wave equation (\ref{eq:1}), the Poynting Theorem becomes,
\begin{equation}
\frac{1}{8\pi}\frac{d}{dt}\left[ \int_{V} \left(\ee E^2+B^2+\frac{\chi}{3\pi}E^3\right) d{\bf r}\right]=-\int_{\partial V} \mathbf{S}\cdot d\mathbf{n}
\label{eq:c1}
\end{equation}
where $V$ is a closed region, and $\partial V$ is its boundary. The vector $\mathbf{S}=\mathbf{E}\times\mathbf{B}$ is the Poynting vector (for simplicity, it is assumed that
$\partial V$ lies in the vacuum  so that
$\mu= \varepsilon =1$,
and $\chi =0$ in a small neighborhood of $\partial V$). In the case of a monochromatic incident wave,
 the flux measured is the time-average
of ${\bf S}$ over a time interval $T\rightarrow\infty$.
By averaging Eq.(\ref{eq:c1}), it then follows that,
\begin{equation*}
\int_{\partial V} \langle \mathbf{S}\rangle\cdot d\mathbf{n}=\mathbf{0},\quad\,\langle \mathbf{S}\rangle=\frac{1}{T}\int_{0}^{T}\mathbf{S}(t)dt,\quad T\rightarrow \infty
\label{eq:c2}
\end{equation*}
This is the flux conservation. In terms of the different harmonics of Eq.(\ref{eq:4}), the time averaged Poynting vector becomes,
\begin{equation*}
\langle \mathbf{S} \rangle=-\frac{c^2}{2\pi\omega}\im\left(\sum_{l=1}^{\infty}\frac{E_{l}\nabla E_{-l}}{l}\right)
\label{eq:c3}
\end{equation*}
Of interest is the flux of the Poynting vector across
the rectangle depicted in Fig.~\ref{fig:1}(b). By Bloch's condition (\ref{eq:5.1}), the contributions to the flux from the faces $L_{\pm 1/2}: x=\pm \frac{1}{2}$ cancel out so that the flux measured is through the vertical faces $L_{\pm}=\{(x,z)|-\frac{1}{2}\leq x\leq \frac{1}{2},\, z\rightarrow\pm\infty\}$. Note that the vanishing of the flux across
the faces $L_{\pm 1/2}$ is a consequence of the fact that the incident
wave is uniformly extended over the whole $x-$axis.
For example, consider the normal incidence ($k_x=0$) with
one diffraction channel open for the incident radiation.
Then the Poynting vector of the reflected and transmitted fundamental harmonic is normal to the structure and, hence,
carries no flux across $L_{\pm 1/2}$. The second harmonic
$(l=2)$
has three forward and three backward scattering channels
open, $m=0,\pm 1$, relative to the $z-$axis.
The wave with $m=0$ propagates in the
direction normal to the structure and does not contribute
to the flux across $L_{\pm 1/2}$. Since the incident
wave has an infinite front along the $x-$axis, so do the scattered waves with $m=\pm 1$. The waves with $m=1$ and
$m=-1$ carry opposite fluxes across each of the faces
$L_{\pm 1/2}$ as the corresponding wave vectors have
the same $z-$components and opposite $x-$components and, hence, the total flux vanishes. For a finite wave front
(but much larger than the structure period), the second
harmonic would carry the energy flux in all the directions
parallel to the corresponding
wave vectors in each open diffraction channel.

If $\sigma_{l}$ is as defined in Section~\ref{sec:4}, then the flux conservation implies that $\sum_{l=1}^{\infty} \sigma_{l}=1$. Therefore, in the perturbation theory used, i.e., when
the system (\ref{eq:7}) is truncated to Eqs.(\ref{eq:11}),
the inequality  $\sigma_{1}+\sigma_{2}\leq 1$
must be verified to justify the validity of the theory.

The conversion ratio $\sigma_{2}$ is
given in Section~\ref{sec:4}.
If only one diffraction channel is open for
the fundamental harmonic, then
the ratio $\sigma_{1}$ of the scattered and incident fluxes of the fundamental harmonic
reads,
\begin{equation*}
\sigma_{1}=|1+T_{0}|^2+|R_{0}|^2
\label{eq:c4}
\end{equation*}
where $T_{0}$ and $R_{0}$ are the transmission and reflection coefficients which are obtained
from the far-field amplitude of $E_{1}$ as,
\begin{equation*}
E_{1}\rightarrow\begin{cases}
\displaystyle{e^{i\rr\cdot\kk}+
R_{0}e^{i\rr\cdot\kk^{-}},\quad\, z\rightarrow -\infty}\\
                              \displaystyle{(1+T_{0})e^{i\rr\cdot\kk},\quad\quad\ \  z\rightarrow +\infty }
                              \end{cases}
                              \label{eq:c5}
                              \end{equation*}
where $\kk=\kx\ex+\kz\ez$ is the incident wave vector
and $\kk^{-}=\kx\ex-\kz\ez$ is the wave vector
of the reflected fundamental harmonic. It then follows from Eqs.(\ref{eq:9}) and (\ref{eq:21}) that
\begin{equation*}
\begin{cases}
\displaystyle{R_{0}=i\frac{2\pi \dl(k)}{\kz}
\Bigl[ (E_{1+}+2\nu E_{2+}\overline{E}_{1+})e^{i h\kz}+(E_{1-}+2\nu E_{2-}\overline{E}_{1-})
e^{-i h\kz}\Bigr]}\\
\displaystyle{T_{0}=i\frac{2\pi \dl(k)}{\kz}
\Bigl[ (E_{1+}+2\nu E_{2+}\overline{E}_{1+})e^{-i h\kz}+(E_{1-}+2\nu E_{2-}\overline{E}_{1-})
e^{i h\kz}\Bigr]}
\end{cases}
\label{eq:c6}
\end{equation*}
In the vicinity of a critical point $(\hb,\kb)$, the coefficients $R_{0}$ and $T_{0}$ obey the estimate,
\begin{equation*}
R_{0}\approx T_{0}\approx i\frac{4\pi \dl(\kb)}{k_{z,b}}\vp E_{1+}\left(1+\frac{\nu^2 |E_{1+}|^2}{\zeta_{b}}\right)
\label{eq:c7}
\end{equation*}
After some algebraic manipulations, it is found that,
\begin{equation*}
\sigma_{1}+\sigma_{2}=1+\frac{8\pi\dl(\kb)}{k_{z,b}}\nu^2|E_{1+}|^4\left[A_{b}+\frac{4\pi\dl(\kb)}{k_{z,b}}\vp^2\left(\re\left\{\frac{1}{\zeta_{b}}\right\}+\frac{\nu^2|E_{1+}|^2}{|\zeta_{b}|^2}\right)\right]
\label{eq:c8}
\end{equation*}
where $A_{b}$ is the constant defined as,
\begin{equation*}
A_{b}=\left[2S_{c}|a+b+1|^2-\im\left\{\frac{1}{\zeta}\right\}\right]_{(h,k)=(\hb,\kb)}
\label{eq:c9}
\end{equation*}
and $S_{c}=\im\{\ald+\bed\}$ is introduced
in Eqs.(\ref{eq:a3}).
Expressing $a$ and $b$ defined by
(\ref{eq:29.1}) via the coefficients $\ald$ and $\bed$ of the symmetric matrix $\hcd$, one also obtains
\begin{equation*}
S_{c}=\im\left\{\frac{a+b}{1+a+b}\right\}
\label{eq:c10}
\end{equation*}
Since at the point $(\hb,\kb)$ the value of $\zeta$ is $\zeta_{b}=(2(a+b))^{-1}|_{(h,k)=(\hb,\kb)}$, it follows that,
\begin{equation*}
A_{b}=\left[2\im\left\{\frac{a+b}{1+a+b}\right\}|a+b+1|^2-2\im\{a+b\}\right]\Bigg|_{(h,k)=(\hb,\kb)}
\label{eq:c11}
\end{equation*}
For general complex numbers $a$ and $b$, the expression in square brackets is always zero. Therefore $A_{b}=0$, and,
\begin{equation}
\sigma_{1}+\sigma_{2}=1+2\left(\frac{4\pi\dl(\kb)}{k_{z,b}}\vp\nu|E_{1+}|^2\right)^2\left(\re\left\{\frac{1}{\zeta_{b}}\right\}+\frac{\nu^2|E_{1+}|^2}{|\zeta_{b}|^2}\right)
\label{eq:c12}
\end{equation}
By Eq.~(\ref{eq:a6}), $\re\{\zeta_{b}^{-1}\}\approx -4$.
In Appendix \ref{asec:3} it is proved that
$\nu^{2}|E_{1+}|^2=O(\vp^{2/3})$. Consequently, near the critical point $(\hb,\kb)$, the right hand summand
in Eq.(\ref{eq:c12}) is negative so that $\sigma_{1}+\sigma_{2}\leq 1$ as required.

\section{Complements on the amplitude $E_{1}$}
\label{asec:3}

The amplitude of the field $E_{1+}$ is a root of the cubic polynomial in Eq.(\ref{eq:30}) which can be solved
by Cardano's method. Put $Y=X+\frac{2}{3}\left(\frac{\vp}{\nu}\right)^2\re\{\zeta\xi\}$. For the new variable $Y$,
Eq.~(\ref{eq:30}) assumes the standard form,
\begin{equation}
Y^{3}+p Y+q=0
\label{eq:b1}
\end{equation}
where,
\begin{equation*}
p=\frac{\vp^4}{3\nu^4}\left(|\zeta\xi|^2-2\re\{(\zeta\xi)^2\}\right),\quad\, q=\frac{2\vp^6}{27\nu^6}\re\{\zeta\xi\}\left(4\re\{(\zeta\xi)^2\}-5|\zeta\xi|^2\right)-\frac{\vp^2}{\nu^{4}}|\zeta|^{4}
\label{eq:b2}
\end{equation*}
As the amplitude $E_{1}$ is uniquely
defined by the system (\ref{eq:7}), it is therefore
expected that the cubic in Eq.(\ref{eq:b1}) should
have a unique real solution in order for the theory to be
consistent. The latter holds if and only if the discriminant
\begin{equation*}
D_{3}=\frac{4}{27}p^3+q^2
\label{eq:b3}
\end{equation*}
 is nonnegative. To prove that $D_3\geq 0$, note first
that $|\zeta\xi|^2-2\re\{\zeta^{2}\xi^{2}\}>0$
in the vicinity of a critical point $(\hb,\kb)$.
This follows from the estimates established
in Appendix~\ref{asec:1}. Indeed, in the
first order of $\delta_{0}(\kb)$,
\begin{equation}
|\zeta_{b}\xi_{b}|^2-2\re\{(\zeta_{b}\xi_{b})^2\}\approx \frac{3\pi^2}{k_{z}^{2}}\delta_{0}^{2}(k)\Bigg|_{k=\kb}>0
\label{eq:b4}
\end{equation}
Next, consider the complex number
\begin{equation*}
\rho=\frac{4}{27}\left[2\re\{\zeta\xi\}\left(2\re\{(\zeta\xi)^2\}-\frac{5}{2}|\zeta\xi|^2\right)+i (|\zeta\xi|^2-2\re\{(\zeta\xi)^2\})^{\frac{3}{2}}\right]
\label{eq:b5}
\end{equation*}
The positivity condition (\ref{eq:b4}) ensures that the coefficient of the complex number $i$ in the expression of $\rho$ is indeed real. After some algebraic manipulations, it can be shown that,
\begin{equation*}
D_{3}=\frac{\vp^4}{\nu^{12}}\Big| |\zeta|^2\nu^2-\vp^4 \rho \Big|^{2}
\label{eq:b6}
\end{equation*}
Thus $D_{3}\geq 0$ as required. The only real solution $Y$ to Eq.(\ref{eq:b1}) is then,
\begin{equation*}
Y=\sqrt[3]{\frac{-q+\sqrt{D_{3}}}{2}}+\sqrt[3]{\frac{-q-\sqrt{D_{3}}}{2}}
\label{eq:b7}
\end{equation*}
It then follows that,
\begin{equation}
|E_{1+}|=\frac{|\vp|^{\frac{1}{3}}}{\nu}\sqrt{\tau_{+}+\tau_{-}},\quad\, \tau_{\pm}=\sqrt[3]{\frac{1}{2}\left(\nu^2|\zeta|^2-\frac{1}{2}\vp^4\re\{\rho\}\pm \Big||\zeta|^2\nu^2-\vp^4\rho\Big|\right)}-\frac{\vp^{\frac{4}{3}}}{3}\re\{\zeta\xi\}
\label{eq:b8}
\end{equation}
provided $\tau_{+}+\tau_{-}\geq 0$. The latter
condition imposes a limit on the validity of the perturbation theory developed in the present study, i.e., the reduction
of the system (\ref{eq:7}) to (\ref{eq:11}) is
justified if $\tau_{+}+\tau_{-}\geq 0$.
This is to be expected because of
the lack of analyticity in $\ch$ of the solution to the nonlinear wave equation (\ref{eq:1}) that can only occur
at the critical points $(\hb,\kb)$ at which bound states in the radiation continuum exist. As one gets away from these critical points in the $(h,k)$-plane, the solution to the nonlinear wave equation becomes analytic in $\ch$, meaning that all the terms that were neglected in finding the principal parts of the amplitudes must now also be taken into account to find a solution befitting the series of Eq.(\ref{eq:2}). The shadowed region depicted in Fig.~\ref{fig:2}(b)
shows the region of the $(h,k)$-plane in which
the condition $\tau_{+}+\tau_{-}\geq 0$ holds for
the first symmetric bound. The presented analysis
of the efficiency of the
second harmonic generation is valid for any choice
of the geometrical parameters, $\Delta h=h-h_b$ and
$R$, and the physical parameters, $\varepsilon_c$ and
$\chi_c>0$, which satisfy the conditions (\ref{eq:39.1})
and $\tau_{+}+\tau_{-}\geq 0$.

\end{document}